\documentclass[12pt]{article} 
\usepackage{amssymb}
\usepackage{amsthm}
\usepackage{amsmath}
\usepackage{graphicx}
\usepackage{graphics}
\usepackage{algorithm}
\usepackage{algorithmic}

\makeatletter
\newcommand{\newalgname}[1]{
  \renewcommand{\ALG@name}{#1}
}

\newalgname{Figure}
\renewcommand{\listalgorithmname}{List the \ALG@name}
\makeatother

\newtheorem {Theorem} {Theorem} [section]
\newtheorem {Lemma}[Theorem] {Lemma}

\title{ Gopala-Hemachandra codes revisited}
\author{ L. Childers  \\ 
         Department of Computer Science \\
         East Carolina University\\
         Greenville, NC - 27858 
       \and
         K. Gopalakrishnan\\
         Department of Computer Science \\
         East Carolina University \\
         Greenville, NC - 27858}

\begin{document} \maketitle

\begin{abstract}

   Gopala-Hemachandra codes are a variation of the Fibonacci universal code
and have applications in cryptography and data compression. We show that
$GH_{a}(n)$ codes always exist for $a=-2,-3$ and $-4$ for any integer $n \geq 1$
and hence are universal codes.
We develop two new algorithms to determine whether a GH code exists for a given
set of parameters $a$ and $n$. In 2010, Basu and Prasad showed experimentally that 
in the range $1 \leq n \leq 100$ and $1 \leq k \leq 16$, there are at most $k$ consecutive integers 
for which $GH_{-(4+k)}(n)$ does not exist. We turn their numerical result 
into a mathematical theorem and show that it is valid well beyond the limited range considered by them. 

{\em keywords:} Fibonacci codes, Zeckendorf representation, Gopala-Hemachandra codes,
Data compression.

\end{abstract}
\vspace {.2in}

\newpage
\section{Introduction}
\label{intro}

    The Fibonacci sequence is a sequence of positive integers whose terms are defined by the recurrence relation $F[n] = F[n-1] + F[n-2]$ for all $n > 2$ with the initial conditions $F[1]=1$ and $F[2]=2$. So, the sequence is $1,2,3,5,8,13, \ldots$. Zeckendorf's theorem \cite{Zeck72} 
states that every positive integer can be represented uniquely as the sum of one or more distinct Fibonacci numbers in such a way that the sum does not include any two consecutive Fibonacci numbers. More precisely, if $n$ is a positive integer, then $n$ can be written uniquely as $\sum_{i=1}^{l} \alpha_{i} F[i]$, where $\alpha_{i}$ is either $0$ or $1$, $\alpha_{l}=1$, if $\alpha_{i}=1$, then $\alpha_{i+1}=0$ and $F[i]$ is the $i^{th}$ Fibonacci number. Such a sum is called the Zeckendorf representation of $n$. The Fibonacci code for $n$, denoted by $F(n)$, is simply the binary string $\alpha_{1}\alpha_{2}\alpha_{3} \ldots \alpha_{l}1$, where a 1 is appended at the end. It is interesting to note that although
the theorem is named after the eponymous author who published his paper in 1972, the same result
had been published 20 years earlier by Gerrit Lekkerkerker \cite {Lek52}.

The Fibonacci code is a universal code which encodes positive integers into binary codewords. It was first defined by Apostolico and Fraenkel \cite{AF87} and has applications in data compression.  The Fibonacci code of any integer has the interesting property that it ends with $11$ and does not have any other consecutive 1's in it. This property makes it a prefix code and thus a uniquely decodable binary code of variable size. Fibonacci coding is a self-synchronizing code, making it easier to recover data form a damaged stream. This robustness property makes it useful in practical applications in comparison to other universal codes. The Fibonacci code for the first $15$ integers are shown in table~\ref{fc}.

\begin{table}[h]
\caption{Fibonacci Code for $1 \leq n \leq 15$}
\label{fc}
\begin{tabular}{||r|c||}
\hline
1 & 11 \\
2 & 011 \\
3 & 0011 \\
4 & 1011 \\
5 & 00011 \\
6 & 10011 \\
7 & 01011 \\
8 & 000011 \\
9 & 100011 \\
10 & 010011 \\
11 & 001011 \\
12 & 101011 \\
13 & 0000011 \\
14 & 1000011 \\
15 & 0100011 \\
\hline
\end{tabular}
\end{table}
 
     The Fibonacci sequence appears in the book {\em Liber Abaci} published by Fibonacci in 1202. However according to \cite{PS85}, the Fibonacci sequence appears in Indian mathematics in connection with Sanskrit prosody apparently as early as 450 B.C. About fifty years before the publication of {\em Liber Abaci}, Gopala and Hemachandra not only independently studied the Fibonacci sequence but also introduced a generalization of Fibonacci sequence known as the Gopala-Hemachandra Sequence (GH sequence for short). The GH sequence is 
defined using the similar recurrence relation $GH[n] = GH[n-1]+ GH[n-2]$ for $n \geq 3$,
with the initial conditions $GH[1]=a$ and $GH[2]=b$, where $a$ and $b$ are arbitrary
integers. In other words, GH sequence is
simply the sequence $a, b, a+b, a+2b, 2a+3b, 3a+5b, \ldots$ where the initial numbers $a$ and $b$ are arbitrary integers. When $a=1$ and $b=2$, it boils down to the Fibonacci sequence.

     J.H. Thomas proposed a variation on the Fibonacci sequence in 2007 \cite{JHT07}. 
It is basically the same as Gopala-Hemachandra sequence in which the integer $a \leq -2$ 
and $b=1-a$. He extended the concept of the ``Zeckendorf representation'' to these variant sequences and also came up with the notion of the Gopala-Hemachandra code for an integer formed in a similar manner to the Fibonacci code. We will denote such a code for an integer $n$ using the notation $GH_{a}(n)$. He also observed that $GH_{a}(n)$ may neither exist nor be unique when it exists. For example, when $a=-5$, he observed that there is no Zeckendorf representation for $5$ or $12$. On the other hand, when $a=-2$, there are two different codes for $GH_{a}(7)$ viz., 01011 and 1000011.

     In a computationally oriented paper \cite{BP10} Basu and Prasad determined $GH_{a}(n)$ codes when they exist, in the range $-2 \leq a \leq -20$ and $1 \leq n \leq 100$. 
They noted that $GH_{a}(n)$ exists for $-4 \leq a \leq -2$ and $1 \leq n \leq 100$.
They also observed $GH_{-(4+k)}(n)$ codes do not exist for at most $k$ consecutive integers where $1 \leq k \leq 16$ and $1 \leq n \leq 100$.

     In this article, we generalize their limited computational observations into general mathematical statements and prove them. In Section~\ref{univ} we show that $GH_{a}(n)$ exists for any positive integer $n$ (not just when $1 \leq n \leq 100$) provided $-4 \leq a \leq -2$. In Section~\ref{alg}, we develop two simple algorithms that determine whether $GH_{a}(n)$ exists for a given combination of parameters $(a,n)$. Finally, in section~\ref{nonexistence}, we show that $GH_{-(4+k)}(n)$ codes do not exist for at most $k$ consecutive integers in general (i.e., without bounding $k$ or $n$ from above). Along the way, we point out some erroneous results published in the literature about Gopala-Hemachandra codes.

\section{Universality of GH codes for $a=-2,-3,-4$}
\label{univ}

    In this section, we prove that $GH_{a}(n)$ exists for any positive integer $n$
provided that $a=-2, -3$ or $-4$. The code is obtained by simply appending a $1$ to the Zeckendorf representation of $n$ and so all we need to do is to show that a Zeckendorf representation exists for any positive integer $n$. It turns out that if an ordinary representation for $n$ exists, then it can be easily transformed into a Zeckendorf representation. We state and prove this result first.

\begin{Lemma}
\label{LL}

    Let there be an ordinary binary representation of $n$ using the integers of the $GH_{a}$ sequence. In other words, let $\sum_{i=1}^{l} \alpha_{i} GH_{a}[i]$, where $\alpha_{i}$ is either $0$ or $1$ and $\alpha_{l}=1$. Then there exists a Zeckendorf representation of $n$. In other words, there exists $\beta_{1}, \beta_{2}, \ldots, \beta_{k}$ such that,  $\sum_{i=1}^{k} \beta_{i} GH_{a}[i]$, where $\beta_{i}$ is either $0$ or $1$, $\beta_{k}=1$ and if $\beta_{i}=1$, then $\beta_{i+1}=0$ (in other words,
no two consecutive numbers will be summed).

\end{Lemma}
~~\\
{\bf Proof:}

     Consider the bit string $\alpha_{1}\alpha_{2}\alpha_{3}\ldots\alpha_{l}$. Scan the string from right to left. Each time we find two consecutive $1$'s, we replace the substring $110$ by the string $001$. It is easy to see that the value of the number represented is preserved as $GH_{a}[i] = GH_{a}[i-1]+GH_{a}[i-2]$. It also follows that at the end of the process, the new bit string obtained $\beta_{1}\beta_{2}\beta_{3}\dots\beta_{k}$ will not have any consecutive 1's in it and so is a Zeckendorf representation of $n$.

     Note that $k$ will either be equal to $l$ or $l+1$. $k$ will be equal to $l+1$ if the rightmost two bits of the $\alpha$ string are 1's and $k$ will be same as $l$ in all other situations. \qed

    We will denote the Gopala-Hemachandra Sequence by $GH_{a}[1],GH_{a}[2], GH_{a}[3], \ldots $.
Recall that $GH_{a}[1] = a, GH_{a}[2] = 1-a$ and $GH_{a}[i] = GH_{a}[i-1]+GH_{a}[i-2]$ for $i \geq 2$.
By the term initial segment of the GH sequence, we will denote the first five elements of it, viz., $GH_{a}[1]$ through $GH_{a}[5]$. We will use the term remaining segment to denote the rest of the GH sequence, i.e, from $GH_{a}[6]$ onwards. Observe that the remaining segment has only positive integers and is a monotonically increasing sequence.

\begin{Theorem}
\label{univ-thm}

    Let $n$ be a positive integer and $a$ be $-2,-3$ or $-4$. Then, there exists a Zeckendorf representation for $n$. Consequently, the Gopala-Hemachandra codes are universal for $a=-2,-3$ or $-4$.
\end{Theorem}
~~\\
{\bf Proof:} \\

    In view of the Lemma~\ref{LL}, it suffices to prove that $n$ is the sum of
some integers in the GH sequence. Let $l$ be the largest index from the remaining segment such that $GH_{a}[l] \leq n$. Note that $l$ is well defined provided $n \geq GH_{a}[6]$. If $n=GH_{a}[l]$, then theorem is trivially true. So, let us assume
that $GH_{a}[l] < n < GH_{a}[l+1]$. Then clearly, $0 < n-GH_{a}[l] < GH_{a}[l+1] - GH_{a}[l] = GH_{a}[l-1] < GH_{a}[l]$. We have picked the integer $GH_{a}[l]$. We now repeat the same process again, but this time using $n^{'} = n - GH_{a}[l]$ as the target integer. Note that the new target integer is smaller than the integer we picked. We repeat the same process again and again, until the target integer becomes smaller than $GH_a[6]$. Let us denote the target integer when we stop as the remainder $r$. Then, clearly $n$ is the sum of all the integers from the GH sequence that we picked plus $r$. We now show that $r$ can be expressed as the sum of some integers from the initial segment of the GH sequence.

    When $a=-2$, the initial segment of the GH sequence is $-2, 3, 1, 4, 5$. In Table~\ref{T-2}, we show how to write $1 \leq r < GH_{2}[6] = 9$ as the sum of some integers from the initial segment. Note that in the table, the second column is a binary vector of size 5, with a $1$ in places corresponding to the GH sequence integer picked.

\begin{table}[h]
\caption{representation of remainders for $a=-2$}
\label{T-2}
\begin{tabular}{||r|c||}
\hline
0 & 00000 \\
1 & 00100 \\
2 & 10010 \\
3 & 10001 \\
4 & 00010 \\
5 & 00001 \\
6 & 00101 \\
7 & 01010 \\
8 & 01001 \\
\hline
\end{tabular}
\end{table}

    When $a=-3$, the initial segment of the GH sequence is $-3, 4, 1, 5, 6$. In Table~\ref{T-3}, we show how to write $1 \leq r < GH_{3}[6]=11$ as the sum of some integers from the initial segment.

\begin{table}[h]
\caption{representation of remainders for $a=-3$}
\label{T-3}
\begin{tabular}{||r|c||}
\hline
0 & 00000 \\
1 & 00100 \\
2 & 10010 \\
3 & 10001 \\
4 & 10101 \\
5 & 00010 \\
6 & 00001 \\
7 & 00101 \\
8 & 10011 \\
9 & 01010 \\
10 & 01001 \\ 
\hline
\end{tabular}
\end{table}

 When $a=-4$, the initial segment of the GH sequence is $-4, 5, 1, 6, 7$. In Table~\ref{T-4} we show to write $1 \leq r < GH_{3}[6]=13$ as the sum of some integers
 from the initial segment.

\begin{table}[h]
\caption{representation of remainders for $a=-4$}
\label{T-4}
\begin{tabular}{||r|c||}
\hline
0 & 00000 \\
1 & 00100 \\
2 & 10010 \\
3 & 10001 \\
4 & 10101 \\
5 & 01000 \\
6 & 00010 \\
7 & 00001 \\
8 & 00101 \\
9 & 10011 \\
10 & 10111 \\
11 & 01010 \\
12 & 01001 \\
\hline
\end{tabular}
\end{table}

  So, provided $a=-2, -3$ or $-4$, the remainder can always represented as the sum of some integers from the initial segment. Thus, provided $a=-2,-3,$ or $-4$, any positive integer $n$ can be represented as a sum of some of the integers in the GH sequence. Now, using the Lemma~\ref{LL} such a representation can be converted to a Zeckendorf representation. Hence, we conclude that the GH codes are universal for $a=-2, -3$ or $-4$. \qed

  In \cite{BD16}, the abovementioned theorem is presented as Theorem 3.2. However, their proof is wrong. For example, when $a=-4$, the code constructed for $n=135$ as per the scheme outlined in their proof results in $10000000111$ which is clearly not a Zeckendorf representation. 

\section{Algorithms to determine existence of GH codes}
\label{alg}

    A natural algorithmic question that arises is to determine whether or not a GH code exists for a given set of parameters $a$ and $n$. In this section, we present two different algorithms to answer this question, prove the correctness of these algorithms and comment on their complexity.

We first prove the following simple arithmetical result. This is similar to the result on Fibonacci numbers found in \cite{L91}.

\begin{Lemma}
\label{sum}

   Let $r$ be an integer greater than or equal to 2. Then,

 \[ \sum_{i=2}^{r} GH_{a}[i] = GH_{a}[r+2] -1. \]
\end{Lemma}

~~\\
{\bf Proof:}

   First recall that, the GH sequence for the parameter $a$ is  \\
$a, 1-a, 1, 2-a, 3-a \dots .........$.

   We will prove this lemma by induction. For the base case, suppose $r=2$.
Then, $\sum_{i=2}^{2} GH_{a}[i] = GH_{a}[2] = 1-a$. Also, $GH_{a}[r+2] -1= GH_{a}[4] - 1 = (2-a) -1 =1-a$. So, the statement is true when $r=2$.

   For the induction step, assume that the statement is true for some integer $r$,
so that we have $\sum_{i=2}^{r} GH_{a}[i] = GH_{a}[r+2] -1$. We need to show that
$\sum_{i=2}^{r+1} GH_{a}[i] = GH_{a}[r+3] -1$.
Now,
\begin{eqnarray*}
\sum_{i=2}^{r+1} GH_{a}[i] & = & \left ( \sum_{i=2}^{r} GH_{a}[i]  \right ) + 
GH_{a}[r+1] \\
          & = & GH_{a}[r+2] - 1 + GH_{a}[r+1] \\
          & = & GH_{a}[r+1] + GH_{a}[r+2] -1 \\
          & = & GH_{a}[r+3] - 1
\end{eqnarray*}   \qed

   To simplify notation, we use $GH[i]$ to denote the $i^{th}$ number in the GH sequence when there is no confusion about the parameter $a$. 
We state and prove the following theorem which forms the basis of our algorithms.

\begin{Theorem}
\label{split}

   Let $n$ be a positive integer. If $n$ can be realized as the sum of some numbers in the GH sequence, then there exists integers $n_{0}$ and $n_{1}$ satisfying the following
conditions
\begin{enumerate}
\item $n = n_{0}+n_{1}$
\item $n_{0} = \displaystyle \sum_{i=1}^{5} \alpha_{i} GH[i]$, where $\alpha_{i}$ are $0$ or $1$ and
$0 \leq n_{0} < GH[6]$.
\item $n_{1} = \displaystyle \sum_{i=6}^{k} \alpha_{i} GH[i]$. where $\alpha_{i}$ are $0$ or $1$, $\alpha_{k} = 1$ and forms the Zeckendorf representation of $n_{1}$ which can be
constructed using greedy algorithm.
\end{enumerate}
\end{Theorem}

~~\\
{\bf Proof:}

      Suppose $n$ can be realized as the sum of some numbers in the GH sequence.
Then, $\displaystyle \sum_{i=1}^{l} \beta_{i} GH[i]$, where $\beta_{i}$ is 0 or 1
and $\beta_{l}=1$. Let $n_{0}^{'} =  \displaystyle \sum_{i=1}^{5} \beta_{i} GH[i]$
and $n_{1}^{'} = \displaystyle \sum_{i=6}^{l} \beta_{i} GH[i]$.
If $0 \leq n_{0}^{'} < GH[6]$, we set $\alpha_{i} = \beta_{i}$ for $1 \leq i \leq 5$. Thus $n_{0} = n_{0}^{'}$ and meets the condition 2 of the theorem.

      Note that if $n_{0}^{'} < 0$, then $\beta_{1} = 1$, $\beta_{2} = \beta_{4} = \beta_{5} = 0$ and $\beta_{3} = 0$ or $1$.
If $n_{0}^{'} < 0$, then $n_{1}^{'} > 0$ as $n$ is a positive integer. Consequently, $\beta_{i} = 1$ for at least one value of $i \geq 6$. Let $j$ denote the smallest index greater than or equal to six, such that $\beta_{j}=1$. Then we could slightly perturb the bits as follows. Set $\beta_{j}=0$, $\beta_{j-1}=1$ and $\beta_{j-2} =1$.
Note that $\beta_{j-1}$ and $\beta_{j-2}$ are guaranteed to be 0 before we changed them. As $GH[j] = GH[j-1] + GH[j-2]$, the sum of the numbers corresponding to 1 bits, still remains the same viz., $n$. We can iterate this process again and again, until $\beta_{5}$ becomes $1$. Note that, when we are through, $\beta_{1}$ through $\beta_{3}$ will remain unchanged and $\beta_{4}$ will either remain 0 or would have changed to 1. Now we set  $\alpha_{i} = \beta_{i}$ for $1 \leq i \leq 5$.
Then clearly $0 \leq n_{0} < GH[6]$

      If $n_{0}^{'} \geq GH[6]$, then $\beta_{4} = \beta_{5} = 1$ or else the first five bits of the $\beta$ sequence is $01101$. In the later case, we can alter it to be
$00011$ without changing $n_{0}^{'}$. So, without loss of generality, we can assume
$\beta_{4}= \beta_{5} = 1$. Now, we can use the same technique used in the proof of Lemma~\ref{LL} to eliminate these two consecutive 1's. Let us denote the resulting bit sequence to be $\alpha_{i}$. Then, clearly $0 \leq n_{0} < GH[6]$, where $n_{0} = 
\sum_{i=1}^{5} \alpha_{i}GH[i]$.

     Now that we have ascertained that $0 \leq n_{0} < GH[6]$, let $n_{1} = n-n_{0}$ so that the condition 1 stated in the theorem is satisfied. We do know that $n_{1}$ is the sum of some numbers in the remaining segment of the GH sequence. If that representation is not Zeckendorf representation, then it can always be transformed into a Zeckendorf representation using Lemma~\ref{LL}.  All that remains to be shown is that the Zeckendorf representation can be found using a Greedy approach.

    In order to construct the Zeckendorf representation of $n_{1}$ using a Greedy approach, we pick the largest GH sequence number not exceeding $n_{1}$ and then iterate the process. Let $GH[m] \leq n_{1} < GH[m+1]$. Then, we claim that must pick $GH[m]$. Suppose we did not pick $GH[m]$. Let $L$ be the largest number that we can form
without picking $GH[m]$. We could pick  the next largest number $GH[m-1]$ and as we cannot pick consecutive numbers from the GH sequence for Zeckendorf representation, we get 

\begin{eqnarray*}
 L  & = & GH[m-1] + GH[m-3]  + \ldots GH[6] \\
    & = & GH[m-2] + GH[m-3] +GH[m-4] + GH[m-5] + \ldots GH[5] + GH[4] \\
    & = & \sum_{i=4}^{m-2} GH[i] \\
    & < & \sum_{i=2}^{m-2} GH[i] \\
    & = & GH[m] -1 \\
    & < & GH[m] \\
    & \leq & n_{1} \\
\end{eqnarray*}

    So, the net result is that $L < n_{1}$. This means we cannot possibly form a Zeckendorf representation of $n_{1}$ using the remaining segment of GH sequence, if we don't make the greedy choice of picking GH[m]. So, we must make the greedy choice of picking $GH[m]$.

After having picked $GH[m]$, we are now trying to make up the number $n_{1}-GH[m]$. It is easy to observe that $0 \leq n_{1} - GH[m] < GH[m-1]$. So, the next integer picked will not be $GH[m-1]$ and so we will not have picked two consecutive GH sequence numbers. After each iteration, our target number decreases and should eventually become zero as we know that there is a Zeckendorf representation of $n_{1}$ using only the remaining segment of the GH sequence. So, we can indeed find the Zeckendorf representation of $n_{1}$ using the Greedy algorithm.  \qed

     We now present our first algorithm to check whether a $GH$ code exists for a given integer $n$. 
\begin{algorithm*}
\caption{Simple Algorithm}
\label{basicalgo}
\begin{algorithmic}[1]

 \FOR {$n_{0} := 1$ \TO $GH[6]-1$}

     \IF {$n_{0}$ can be represented using the initial segment of the GH sequence}

         \STATE $n_{1} = n - n_{0}$
         \IF {Zeckendorf representation of $n_{1}$ using the remainder segment of the GH sequence can be found by applying the
Greedy technique}
         \STATE $GH$ code for $n$ exists. 
         \STATE Concatenate the representation of $n_{0}$ and $n_{1}$.
         \STATE Use the technique of Lemma~\ref{LL} to make it a Zeckendorf representation, if it is not already a Zeckendorf representation.

         \STATE Print code for $n$
         \STATE Exit

         \ENDIF

      \ENDIF

  \ENDFOR

  \STATE $GH$ Code for $n$ does not exist.

\end{algorithmic}
\end{algorithm*}

   The correctness of the algorithm shown in Figure~\ref{basicalgo} follows immediately from the proof of Theorem~\ref{split}. As the initial segment is fixed in size, the representation of $n_{0}$ can be found in constant time, if it exists. The greedy technique to find a Zeckendorf representation for $n_{1}$, if it exists, runs in linear time (proportional to length of representation of $n_{1}$ and thus of $n$ as $n \geq n_{1}$). Also,
 it takes only linear time to perform line no. 7 which is applying the technique of Lemma~\ref{LL}. 
As $GH[6]$ is completely determined by the parameter $a$, the number of times the for loop is run is a constant (i.e., independent of $n$). So, the entire algorithm runs in linear time, when $a$ is considered a fixed parameter and $n$ is considered the input.

   In practice, we will use the above algorithm only when $a < -4$, as we have already shown in Section~\ref{univ} that GH code exists for any positive integer $n$, when $a=-2, -3$ or $-4$. Therefore, let us assume that $a=-(4+k)$. Note that when $a=-(4+k)$, the GH sequence is $-(4+k), k+5, 1, k+6, k+7, 2k+13 \dots$. It is not difficult to
see that, when $a=-(4+k)$, there are exactly 13 integers $n_{0}$ such that
$0 \leq n_{0} \leq GH[6]$ for which there is a representation using the initial segment of the GH sequence. Those thirteen integers along with their representation are given in Table~\ref{T-k}.

\begin{table}
\caption{representation of remainders for $a=-(4+k)$}
\label{T-k}
\begin{tabular}{||c|c||}
\hline
0 & 00000 \\
1 & 00100 \\
2 & 10010 \\
3 & 10001 \\
4 & 10101 \\
k+5 & 01000 \\
k+6 & 00010 \\
k+7 & 00001 \\
k+8 & 00101 \\
k+9 & 10011 \\
k+10 & 10111 \\
2k+11 & 01010 \\
2k+12 & 01001 \\
\hline
\end{tabular}
\end{table}

   So, representation for $n_{0}$ using the initial segment, does not exist if $5 \leq n_{0} \leq k+4$ or if $k+11 \leq n_{0} \leq 2k+10$. In all other cases, where $0 \leq n_{0} \leq GH[6]$, representation for $n_{0}$ and can be found by looking up in the Table~\ref{T-k}. As such, line 2 of the algorithm shown in Figure~\ref{basicalgo} can be implemented in the above manner.

   We now develop a more efficient algorithm than the Algorithm shown in Figure~\ref{basicalgo}. Our efficient algorithm is depicted in Figure~\ref{effalgo}. The algorithm relies crucially on the following two lemmas.

\begin{algorithm*}
\caption{Faster Algorithm}
\label{effalgo}
\begin{algorithmic}[1]

       \STATE Apply Greedy Technique to represent $n$ using only the remaining segment of the GH sequence.

       \STATE Let $n_{1}$ be the sum of the numbers picked up.

       \STATE $n_{0} = n - n_{1}$

       \IF {$ (5 \leq n_{0} \leq k+4)$ or $(11+k \leq n_{0} \leq 10+2k) $}
         \STATE $n_{0} = GH[2] + GH[4]$.
         \STATE $n_{1} = n - n_{0}$ 
         \STATE Attempt to find Zeckendorf representation of $n_{1}$ by applying the greedy technique to the remaining segment.
         \IF {it exists}
         \STATE Concatenate the representation of $n_{1}$ to ``01010".
         \STATE Print the result. 
         \ELSE  
         \STATE Print code for $n$ does not exist.
       \ENDIF

       \ELSE
         \STATE Find representation of $n_{0}$ by looking up in Table~\ref{T-k}
         \STATE Concatenate the representation $n_{0}$ and the representation of $n_{1}$
         \STATE Use the technique of Lemma~\ref{LL} to make it a Zeckendorf representation, if it is not already a Zeckendorf representation.
         \STATE Print the result 

      \ENDIF

\end{algorithmic}
\end{algorithm*}

\begin{Lemma}
\label{ng2}

   Let $n$ be a positive integer. Suppose a GH code exists for $n$ that is not produced by the greedy algorithm, then the second bit of the code must be a $1$.

\end{Lemma}

~~\\
{\bf Proof:}

   Recall that the greedy algorithm proceeds by repeatedly picking the largest GH sequence number that can be picked while making sure that the sum of the numbers picked does not exceed $n$. If the GH code for $n$ is not produced by the greedy algorithm, then it picks a different number than suggested by the greedy algorithm at some point. Let us focus on the first time there is a difference. Let $r$ denote the difference between $n$ and the numbers already picked up to this point of time. Suppose the greedy algorithm suggests picking $GH[i]$ as $i$ is the largest index such that $GH[i] \leq r$. If we don't pick $GH[i]$, let $L$ denote the largest number that we can form using the other GH sequence numbers while not including GH[2]. Then, we have

\begin{eqnarray*}
L & = & GH[i-1] + GH[i-3] + \ldots + GH[4] \\
  & = & GH[i-2] + GH[i-3] + GH[i-4] + GH[i-5] + \ldots + GH[3] + GH[2] \\
  & = & \sum_{j=2}^{i-2} GH[j] \\
  & = & GH[i] - 1 \\
  & < & GH[i] \\
  & \leq & r  \\
\end{eqnarray*}

  So, it is impossible to make up the remainder $r$, unless $GH[2]$ is also used.
Hence, the second bit of the GH code for $n$ must be a $1$. \qed.

\begin{Lemma}
\label{ng4}

   Let $n$ be a positive integer. Suppose a GH code exists for $n$ that is not produced by the greedy algorithm, then the fourth bit of the code must be a $1$.

\end{Lemma}

~~\\
{\bf Proof:}

   By Lemma~\ref{ng2}, we already know that the second bit of the GH code must be a $1$. So, the third bit must be a zero, as the code cannot have two consecutive 1's in it.

If the GH code for $n$ is not produced by the greedy algorithm, then it picks a different number than suggested by the greedy algorithm at some point. Let us focus on the first time there is a difference. Let $r$ denote the difference between $n$ and the numbers already picked up to this point of time. Suppose the greedy algorithm suggests picking $GH[i]$ as $i$ is the largest index such that $GH[i] \leq r$. If we don't pick $GH[i]$, let $L$ denote the largest number that we can form using the other GH sequence numbers while not including GH[4]. Then, we have
\begin{eqnarray*}
L & = & GH[i-1] + GH[i-3] + \ldots + GH[6] + GH[2] \\
  & = & GH[i-2] + GH[i-3] + GH[i-4] + GH[i-5] + \ldots + GH[5] + GH[4] + GH[2]\\
  & = & \sum_{j=4}^{i-2} GH[j] + GH[2] \\
  & < & \sum_{j=2}^{i-2} GH[j] \\
  & = & GH[i] - 1 \\
  & < & GH[i] \\
  & \leq & r \\
\end{eqnarray*}

 So, it is impossible to make up the remainder $r$, unless $GH[4]$ is also used.
Hence, the fourth bit of the GH code for $n$ must be a $1$. \qed.

 Given an integer $n$, we can use the greedy algorithm over the remaining segment  of
the GH sequence (i.e., from GH[6] onwards) and let $n_{1}$ denote the sum of the numbers from the sequence picked up. Let $n_{0} = n-n_{1}$. If $n_{0}$ has a representation using the initial segment of the sequence (which can be determined by looking up the Table~\ref{T-k}), then we can concatenate the code for $n_{0}$ and the code for
$n_{1}$ and then apply Lemma~\ref{LL} if needed to get the code for $n$.

 Note that if $n_{0}$ cannot be represented using the initial segment, there is still a possibility that GH code for $n$ exists. However, in this case the code must be something that is not produced by the greedy algorithm working with input $n$.

 In view of Lemma~\ref{ng2} and Lemma~\ref{ng4}, we know that if there is a GH code for $n$, that is not produced by the greedy algorithm, then the second bit and the fourth bit must be $1$. However, as the GH code cannot have consecutive 1's except at the end, the initial segment of the GH code must be ``01010" (if the code length is longer than 5). Now, let $n_{0} =
GH[2] + GH[4] $. Let $n_{1} = n - n_{0}$. In view of Theorem~\ref{split}, we can claim that $n_{1}$ must be greedily constructible using the remaining segment of the GH sequence, as this is the only value for $n_{0}$ that works for this $n$. This establishes the correctness of algorithm shown in Figure~\ref{effalgo}.

  Both the algorithms have the same asymptotic complexity, viz., linear time in the length of the representation of code for $n$. However, the second algorithm is much more efficient in a practical sense, as it constructs the code for $n$ in just two attempts, if it exists.

  In \cite{PD19}, the authors present two algorithms to determine whether $GH$ code exists for a given set of parameters $a$ and $n$. Their second algorithm makes use of their first algorithm and so we will focus on their first algorithm only here. The main issue is that although they have presented the algorithm (see Method 1 in page 164), they did not bother to prove that their algorithm is indeed correct and also they did not bother to analyze the complexity of their algorithm. Also, the algorithm is not described precisely and hence is hard to follow. Finally, their algorithm as presented is wrong. For example, if $a=-6$, the algorithm returns the string ``10000000110011" as the code for $n=649$, which is clearly wrong as it contains consecutive 1's in the middle.

\section{Non-existence of GH codes for consecutive integers}
\label{nonexistence} 

     In \cite{BP10}, Basu and Prasad observed that there are at most $k$ consecutive integers for which $GH_{-(4+k)}$ code does not exist when $1 \leq k \leq 16$ and $1 \leq n \leq 100$. They made this observation using the tables they created for GH codes when $-20 \leq a \leq -2$ and $1 \leq n \leq 100$. 
We turn their numerical result 
to a mathematical theorem and show that the result is valid well beyond the limited range considered by them. We show that the result is true, even when $k > 16$. Moreover, we show that the result is true for the entire set of positive integers (i.e., there is no need to upper bound $n$).

\begin{Theorem}
\label{k-cons}

    Let $k$ be a positive integer. Then, there exists at most $k$ consecutive integers for which $GH_{-(4+k)}$ codes do not exist. 

\end{Theorem}

~~\\
{\bf Proof:}

    Let $n$ be an integer for which a $GH_{-(4+k)}$ code does not exist.
Let us attempt to construct $n$ using the greedy algorithm applied to the
remaining segment of the $GH$ sequence.
In this process, we would have picked some integers from the remaining segment of
the GH sequence. Let the sum of the integers we picked be $n_{1}$. Let $n_{0} = n - n_{1}$. Observe that
$n_{0} < GH[6]$, as otherwise, we would have picked $GH[6]$ in the last step of
the greedy algorithm. Then, it should not be possible to represent $n_{0}$ as the sum of some integers from the initial segment of the GH sequence (i.e., from GH[1] to GH[5]). For, if we were able to represent $n_{0}$ as the sum of some integers from the initial segment, then by Theorem~\ref{split}, we know that there is a GH code for $n$ which contradicts with the assumption made at the beginning.

    Now, let  $n$ and $n^{'}$ be two consecutive integers (with $n < n^{'}$) for which $GH_{-(4+k)}$ codes do not exist. When we run the greedy algorithm on $n$ and $n^{'}$, we will choose the same number from the GH sequence at every step. Hence $n_{1}$ will be same as $n^{'}_{1}$. So, we can conclude that $n_{0}$ and $n^{'}_{0}$ will be two consecutive integers, that are not representable as the sum of some integers from the initial segment. Moreover, $0 \leq n_{0}, n^{'}_{0}  < GH[6]$.

    By extension of the above argument, we can claim that if there are $k+1$ consecutive integers for which $GH_{-(4+k)}$ code does not exist, then there must be $k+1$ consecutive integers (each one of which is greater than or equal to $0$ and less than $GH[6]$), that are not representable as the sum of some integers from the initial segment of the GH sequence.
However, as we can see from Table~\ref{T-k}, this is not possible. The sequence of numbers from $5$ to $k+4$ are not representable, but this sequence has only $k$ elements. Again, the sequence of numbers from $k+11$ to $2k+10$ are not representable, but this sequence also has only $k$ elements.

    Thus, there does not exist $k+1$ consecutive integers for which $GH_{-(4+k)}$ code does not exist. \qed

\section{Concluding Remarks}
\label{con}

     In this article, we showed that a $GH$ code exists for any positive integer $n$, when $a=-2,-3$ and $-4$. We presented two algorithms to determine whether or not a $GH$ code exists for a given set of parameters $a$ and $n$. Moreover, our algorithms construct such a code, if it exists. Both our algorithms run in linear time in the length of the code, but our second algorithm is faster than the first in practice. Finally, we proved that there are at most $k$ consecutive integers for which a $GH_{-(4+k)}$ code does not exist. This result is observed to be true earlier in limited ranges for $n$ and $k$. But, we are able to establish it unconditionally.

     The third order Gopala-Hemachandra codes $GH_{a}^{3}$ are defined using the recurrence relation $GH[n] = GH[n-1] + GH[n-2] + GH[n-3]$, for $n \geq 4$ and the initial conditions $GH[1] = a$,
$GH[2] = 1-a$ and $GH[3] = 1$. In \cite{NO15}, Nalli and Ozyilmaz determined tables of $GH_{a}^{3}$
codes for $-20 \leq a \leq -2$ and $1 \leq n \leq 100$. However, \cite{NO15} is a computationally oriented paper and does not have general results. It appears that $GH_{a}^{3}$ codes are universal for $-10 \leq a \leq -2$. Moreover, it appears that there exists at most $k$ consecutive integers for which $GH_{-(10+k)}^{3}$ code does not exist. Both of these observations are true in the limited range for which tables are constructed in \cite{NO15}. It would be nice to prove these two statements. Also, it would be nice to come up with an algorithm that, given $a$ and $n$, determines whether $GH_{a}^{3}$ code exists for $n$ and if so, constructs the code.

     In a natural way, higher order Gopala-Hemachandra codes (fourth order, fifth order etc) can be defined. It would be nice to come up with a function $f(m)$ and prove that $GH_{a}^{m}$ codes are universal if and only if $ -f(m) \leq a \leq -2$. Finally, it would also be interesting to see whether a very general algorithm can be developed that takes $a,m$ and $n$ as input and determines whether $GH_{a}^{m}$ code exists for integer $n$ and if so, constructs the code.

\bibliographystyle{plain}

\end {document}